\documentclass[%
 reprint, superscriptaddress,
 amsmath,amssymb,
 aps,
]{revtex4-2}

\usepackage{graphicx}
\usepackage{dcolumn}
\usepackage{bm}
\usepackage{float}
\usepackage{color}
\usepackage{enumitem}

\begin{document}

\preprint{APS/123-QED}

\title{Motility-induced shear thickening in dense colloidal suspensions}

\author{A. Gülce Bayram}\email{gulce.bayram@bilkent.edu.tr}
\affiliation{Department of Mechanical Engineering, Bilkent University, Cankaya, Ankara 06800, Turkey}

\author{Fabian Jan Schwarzendahl}
\affiliation{Institut f\"ur Theoretische Physik II: Weiche Materie, Heinrich-Heine-Universit\"at D\"usseldorf, 40225 D\"usseldorf, Germany}

\author{Hartmut L\"owen}%
\affiliation{Institut f\"ur Theoretische Physik II: Weiche Materie, Heinrich-Heine-Universit\"at D\"usseldorf, 40225 D\"usseldorf, Germany}

\author{Luca Biancofiore}
\affiliation{Department of Mechanical Engineering, Bilkent University, Cankaya, Ankara 06800, Turkey}

\date{\today}

\begin{abstract}
Phase transitions and collective dynamics of active colloidal suspensions are fascinating topics in soft matter physics, particularly for out-of-equilibrium systems, which can lead to rich rheological behaviours in the presence of steady shear flow. In this article, the role of self-propulsion in the rheological response of a dense colloidal suspension is investigated by using particle-resolved simulations. First, the interplay between activity and shear in the solid to the liquid transition of the suspension is analysed. While both self-propulsion and shear destroy order and melt the system by themselves above their critical values, self-propulsion lowers the stress barrier that needs to be overcome during the transition. Once the suspension reaches a non-equilibrium steady state the rheological response is analysed. While passive suspensions show a solid-like behaviour, turning on particle motility fluidises the system
and, at low self-propulsion, the suspension behaves as a shear-thinning fluid. Increasing the self-propulsion of the colloids induces a transition from a shear-thinning to a shear-thickening behaviour, which we attribute to clustering in the suspensions induced by motility. This interesting phenomenon of motility-induced shear thickening (MIST) can be used to tailor the rheological response of colloidal suspensions.
\end{abstract}

\maketitle

\section{Introduction}

During the last decade, active matter has become a topic of intense research~\cite{marchetti2013hydrodynamics,ramaswamyactive}. In particular, 
active colloids have been investigated~\cite{bechinger2016active} since they provide a well-controlled testing ground for out-of-equilibrium systems. Experimentally, one among many realizations of active particles is active Janus colloids~\cite{bechinger2016active}, which can show fascinating phenomena such as motility-induced phase separation~\cite{buttinoni2013dynamical,palacci2013living}, vortex formation~\cite{bricard2015emergent}, clustering induced by hydrodynamic fluxes~\cite{mousavi2019clustering} or wall accumulation~\cite{narinder2019active,volpe2011microswimmers,maggi2016self}.
Dense suspensions of active particles have been realized and studied experimentally for Janus colloids\cite{klongvessa2019nonmonotonic}, for vibrated active disks\cite{BriandPRL2018,BriandPRL2016} and theoretically~\cite{bialke2012crystallization,henkes2011active,de2018static}. Active particles that are in a glassy state~\cite{janssen2019active} have given insights into random close packing~\cite{ni2013pushing} and it has been shown that shearing an active glass former leads to ordering~\cite{mandal2021shear}. However, the rheological properties of dense active colloids are largely unknown.

\begin{figure}[h!]
    \centering
    \includegraphics[width=0.48\textwidth]{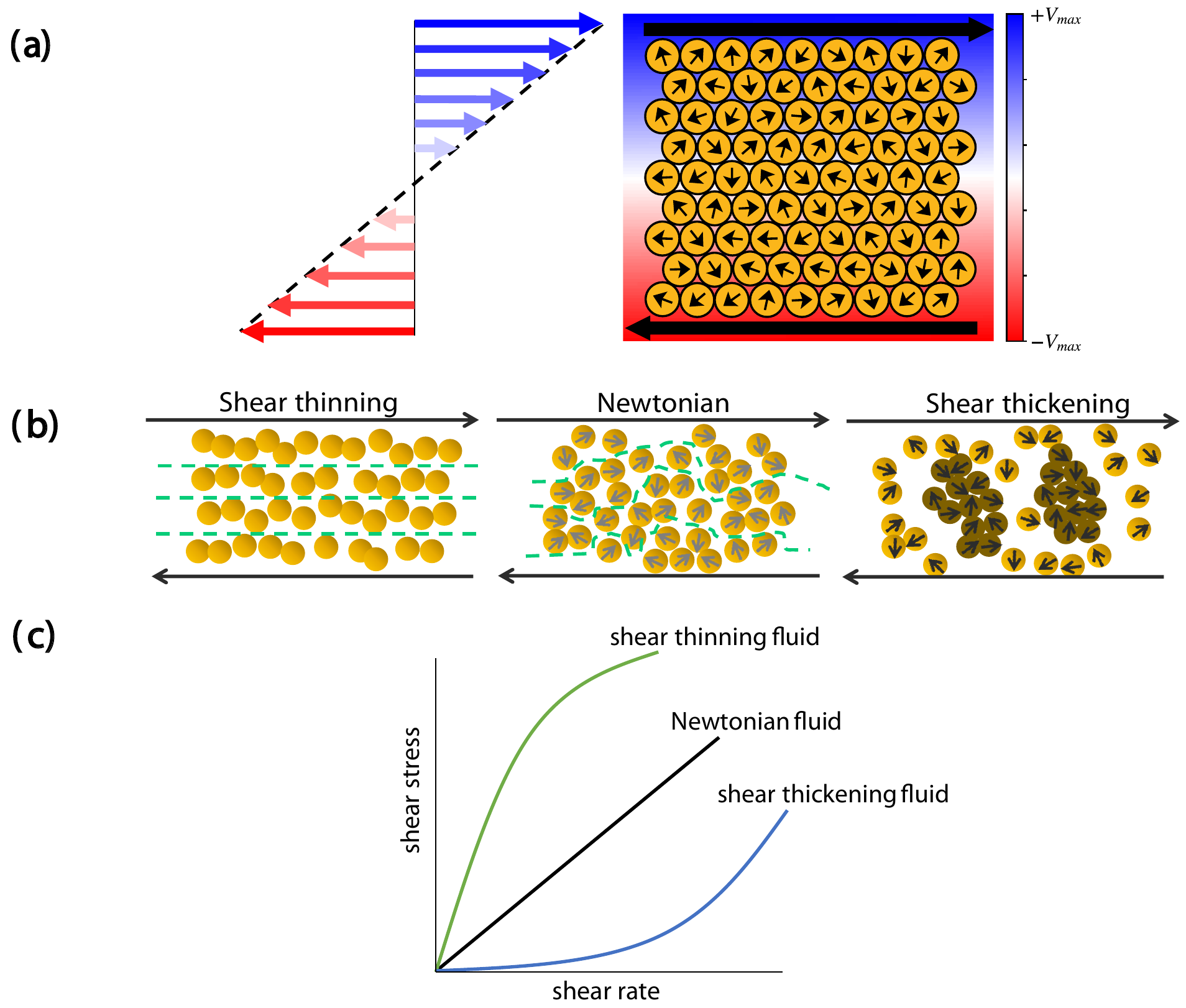}
    \caption{Schematic representations of (a) the sheared active colloidal suspensions with Lees-Edwards boundary conditions (the colour bar represents the distribution of shear force throughout the computational box), (b) the expected particle arrangements in different fluids, (c) shear stress-shear rate curve for different fluids. }
    \label{fig:Schematic}
\end{figure}

On the side of biological microswimmers, such as bacteria or microalgae, it has been shown that the presence of a small fraction of active swimmers in a fluid medium can fundamentally change the fluid's rheological properties. It was found experimentally that pusher-type swimmers such as the bacterium \textit{Escherichia coli} reduce the effective viscosity~\cite{sokolov2009reduction,gachelin2013non,lopez2015turning,liu2019rheology,martinez2020combined} while puller-type swimmers such as the microalgae \textit{Chlamydomonas reinhardtii} can increase the viscosity~\cite{rafai2010effective}. This response to the swimming behavior has also been explained in theoretical and numerical studies~\cite{ishikawa2007rheology,haines2009three,saintillan2010dilute,ryan2011viscosity,moradi2015rheological,bechtel2017linear,hayano2022hydrodynamic,cates2008shearing}. Further, the rheological properties of active fluids, that can model driven microtubules or active acting filaments, have been computed using field theoretical approaches~\cite{hatwalne2004rheology,giomi2010sheared,cates2008shearing,loisy2019exact}.

Here, we investigate the rheological properties of dense active colloids in two spatial dimensions using Brownian dynamics computer simulations. Investigating the strain-stress curves shows that activity reduces and even destroys the stress barrier that a shear flow has to overcome to fluidise the system. 
In the steady state, the system's shear stress reveals that particle motility fundamentally changes the rheological properties of the system: at none or low activity the system is shear thinning, for intermediate activities it becomes Newtonian and at very high activity it is shear thickening (Fig.~\ref{fig:Schematic}). The shear thickening behaviour is induced by particle clusters that stem from the active motion of the particles. Therefore we refer to this new phenomenon as {\it motility-induced shear thickening} (MIST).
In fact, MIST is somehow a consequence of motility-induced phase separation (MIPS)~\cite{buttinoni2013dynamical,palacci2013living} in the bulk which shows a pre-clustering in the one-fluid phase even before full phase separation is reached. These clusters are responsible for the shear-thickening under shear. The full rheological response is further well characterized using a power-law model for the stress as a function of the shear rate, which shows the continuous transition from shear thinning to shear thickening as the activity is increased. Therefore the rheological behaviour can be tuned by activity.

\section{Simulation method}
We study a suspension of $N$ self-propelled particles moving in two spatial dimensions under shear flow, Fig.\ref{fig:Schematic}(a). In section \ref{section21}, we introduce the model to be used for simulating this system, including the units and parameters used in it.  We then describe in section \ref{section22} the observables that we measure in our simulation in order to analyse  (i) the phase transition behaviour and (ii) the rheological response of the system across the different self-propulsion and shear forces.    

\subsection{Model}\label{section21}
 In two spatial dimensions, the over-damped dynamics of the colloids are modeled by active Brownian particles in the presence of a steady shear rate $\dot \gamma$, 
\begin{align}
 &\frac{d {{\mathbf{r}}_i}}{d t} = - \frac{1}{\Gamma} \sum_{i<j}{{\mathbf{\nabla}}_i} U_{i j} + v_0 \mathbf{e}_{i} + \mathbf{\xi}_{i} + \dot \gamma y_i(t) \mathbf{e}_{x}, \label{eq:shearapb} 
\\
&\frac{d \theta_{i}}{d t}= \eta_{i}(t) + \dot \gamma /2.
\end{align}
The self-propulsion (or motility) has a constant magnitude of $v_0$ in a direction of a unit vector $\bm{e}_{i}= (\mathrm{cos}
\theta_{i} (t), \mathrm{sin} \theta_{i} (t)) $ for particle \textit{i}. The orientation angle $\theta_i$  changes diffusively through $\eta_i$, which is a Gaussian white noise with zero mean $ \bigl \langle \eta_{i} (t)  \bigr \rangle = 0 $ and variance $ \bigl \langle \eta_{i}(t) \eta_{i}(t')  \bigr \rangle = 2 D_r \delta_{i j} \delta(t-t')$. The rotational diffusion coefficient $D_r$ controls the strength of the rotational fluctuations.
The colloidal particles interact via a repulsive Yukawa potential\cite{lowen1992structure}
\begin{equation}
    U(r_{ij})= U_0 a \frac{\mathrm{exp}(- \lambda r_{ij}/a)}{ r_{ij}}, 
\end{equation}
where  $\lambda$ is the screening parameter, $r_{i j}=|r_i -r_j|$ is the distance between two particle centers, $a$ is a typical length scale taken as a unit of length and $U_0$ is the bare potential strength. The Gaussian white noise, $\xi_i$ mimics the stochastic interactions with a thermal bath, with zero mean and variance $ \bigl \langle \xi_{i}(t) \xi
_{i}(t')  \bigr \rangle = 2 D_0 \delta_{i j} \delta_{x y} \delta(t-t')$, where $D_0$ is the translational diffusion coefficient. The last term in Eq.\eqref{eq:shearapb} represents the imposed shear in the $x$ direction ($\bm{e}_x$ is the unit vector pointing into the $x$ direction) with respect to the position of particle $i$ along the shear gradient direction $y$. Finally, $\Gamma $ is the friction coefficient.

 We perform steady shear simulations for different number densities $\rho_N$, the ratio of number of particles and the system area, with Lees-Edwards periodic boundary conditions\cite{lees1972computer}, schematically shown in Fig.\ref{fig:Schematic}(a). The dimensions of the computational box are chosen as $L_x,L_y=(30a,26a)$. Three different numbers of particles $N=900$, $648$ and $528$ are used in the simulations corresponding to three different reduced number densities of $\rho_N a^2 = 0.68, 0.83, 1.15$.
 The  bare potential strength is fixed to $U_0 = 800 k_B T$ and the simulations are started from a hexagonal crystal configuration of the particles. Further a soft potential with the screening parameter $\lambda =3.5$ is chosen. The time step for the evolution of the particle positions is $\Delta t =0.0001 \tau_{0}$, with the natural unit of time $\tau_{0}=a^2/D_0$. The colloid-colloid pairwise interactions are cut off at an inter-particle distance of $2a$ and the rotational diffusion coefficient is $D_r= 3.5 \tau^{-1}$. After an initial relaxation time of $t=0.4\tau_0$, shear is imposed.

 \begin{figure*}[!ht]
    \centering
    \includegraphics[width=1\textwidth]{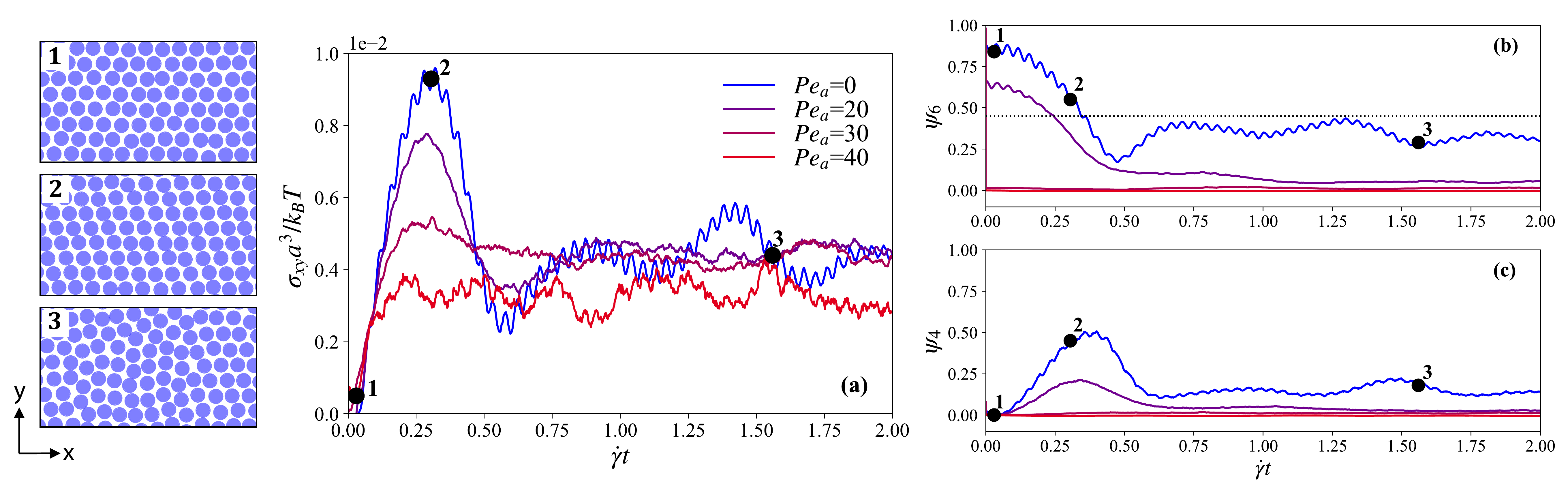}
    \caption{(left) Snapshots of particle configurations for (1) solid-like, (2) cubic-like and (3) liquid-like phases of the passive system, in which the system conditions correspond to the numbered black dots in the plots: (a) The instantaneous shear stress ($\sigma_{xy}$) with respect to strain ($\gamma = \dot{\gamma}t$)  for passive ($Pe_a =0$) and active systems ($Pe_a >0$). The corresponding structural changes throughout the bond orientational order parameters, $\psi_6$ (b) and $\psi_4$ (c). The black dotted-horizontal line defines the position of structural transition for the hexagonal configuration, $\psi_6 \approx 0.45$\cite{bialke2012crystallization}.}
    \label{fig:stress_peak}
\end{figure*}  

\subsection{Measured quantities}\label{section22}
The shear stress, which is the off-diagonal component of the stress tensor, is measured by 
\begin{equation} 
\sigma_{x y} = - \left\langle\frac{1}{2\Omega N} \sum_{i=1}^{N} \sum_{i>j} \frac{\partial U}{\partial r_{i j}} \frac{r_{i j}^x r_{i j}^y}{{r_{i j}}}\right\rangle,
\end{equation}
which is averaging out the stress contribution of every particle in the steady state time frame. $r_{i j}^x$ and $r_{i j}^y$ are the distance between the centers of two particles $i$ and $j$ in direction of $x$ and $y$, respectively.  The stress is nondimensionalised by considering the thermal energy $k_B T$ and length scale $a$ as $\sigma_{x y} a^3 / k_B T$, where $k_B$ is the Boltzmann constant and temperature $T$ with $k_B T=D_0 \Gamma$. 
The dynamics of the system is controlled by two dimensionless numbers.\begin{enumerate}[label=(\roman*)]
\item First, the imposed shear rate is given by {\it the shear Peclet number}, $ Pe_s= \dot{\gamma} a^2 / D_0 $, which measures the ratio of the shear force to thermal fluctuations. \item The second dimensionless number is {\it the active Peclet number}, $ Pe_a=  v_0 a/ D_0  $, which measures the ratio between the self-propulsion force (or the motility strength) and thermal fluctuations. 
\end{enumerate}The typical magnitude of the dimensionless numbers considered here are $Pe_a \in [0,150]$ for the active Peclet number and $Pe_s \in [0,40] $ for the shear Peclet number.

The structural changes in the system are monitored by the 2D bond orientational order parameters\cite{steinhardt1983bond},
\begin{equation}
\psi_{\nu} = \Biggl \langle \Big| \frac{1}{N} \sum_{i=1}^{N} \frac{1}{{\nu}}\sum_{j \in \mathcal{N} (i) } e^{i{\nu} \theta_{i j}}  \Big|^{2} \Biggr \rangle  \ 
\end{equation}
where $\mathcal{N} (i)$ is the set of $\nu$
nearest neighbour particles of the $i$th particle and $\theta_{i j}$ is the angle between the bond vector pointing from particle $i$ to $j$ and horizontal fixed axis. For $\nu=6$ the order parameter is the hexagonal order parameter, which gives zero in the disordered phase whereas it equals 1 in a perfect hexagonal crystal. Similarly to the hexatic order parameter $\psi_6$, we also analyse the cubic orientational  order by using $\nu=4$.
\begin{figure*}[!ht]
    \centering
    \includegraphics[width=1\textwidth]{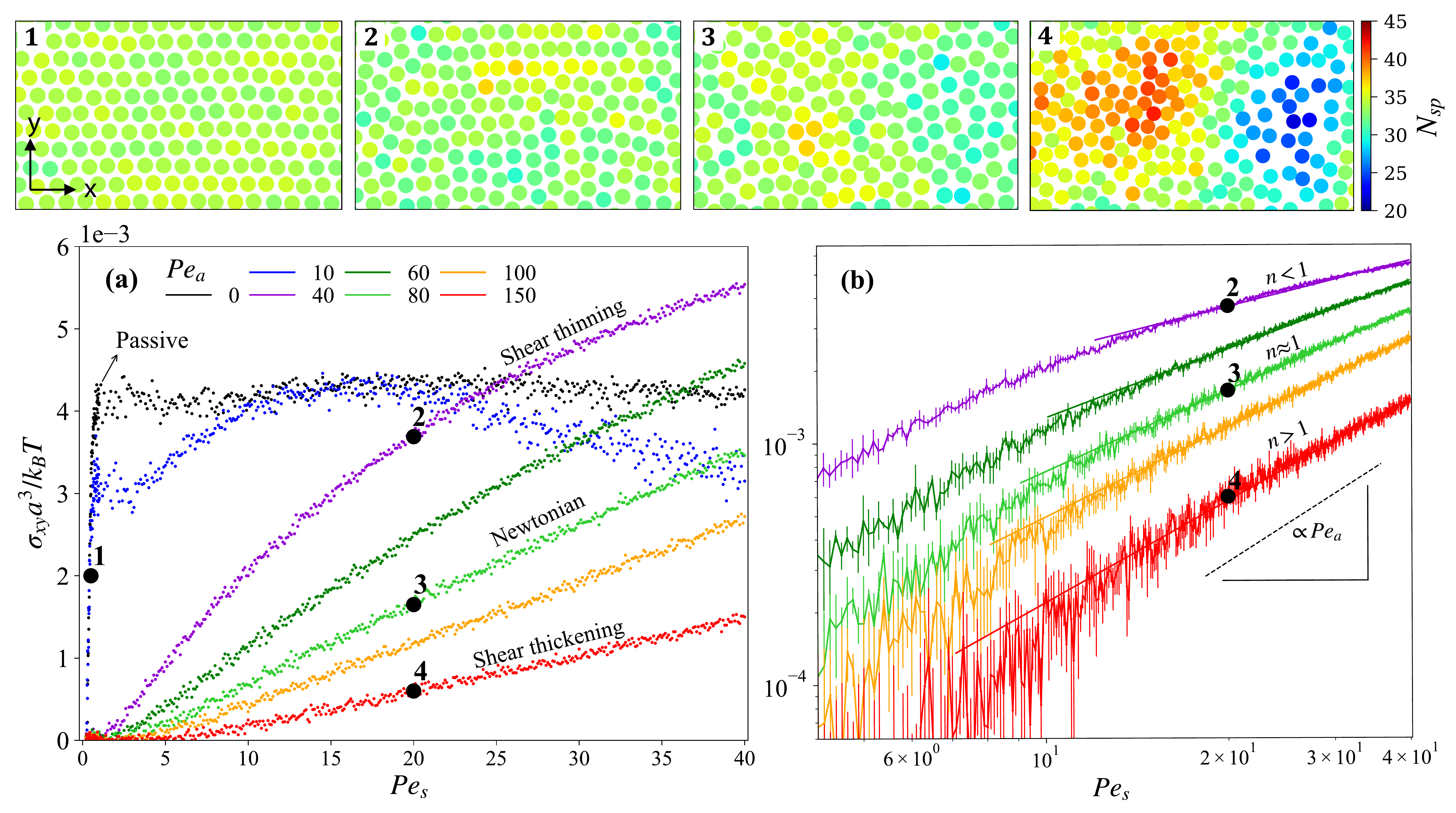}
    \caption{ (top) Snapshots of particle configurations: (1) solid-like, (2) shear-thinning, (3) Newtonian and (4) shear-thickening, where the particles are coloured with respect to their number of surrounding particles ($N_{sp}$). (a) Nondimensional shear stress $ \bigl \langle \sigma_{xy} \bigr \rangle $ - shear rate $\dot{\gamma}$ curves for different self-propulsion forces. (b) Liquid-like regime shown in log-log scale where the power law fitting has been applied is indicated by the fitting lines for different rheological behaviours with respect to the corresponding power indexes($n$). The numbered black dots represent the conditions of the simulations where the snapshots are taken.} \label{fig:power_law}
\end{figure*}
\section{Results}
\subsection{Shear and activity induced melting}\label{section31}
Our particular interest is the characterization of self-propelled particles in the dense regime, $\rho_N a^2 = 1.15$.  We first analyze the shear stress ($\sigma_{xy}$) - strain ($\gamma = \dot{\gamma}t$) relation of the system at a constant shear rate $Pe_s =20 $ for different self-propulsions. Meanwhile, we also monitor the bond-orientational order parameters of the system. We note that the results represented here show the system behaviour starting from the time we impose the shearing force $(t>0.4\tau_0)$. 

In the absence of particle motility ($Pe_a =0$), the system starting out of the perfect hexagonal crystal stays in this solid-like phase during the relaxation, as in Fig.\ref{fig:stress_peak} snapshot(1). When shearing is turned on thereafter, shear stress starts building up to a peak value, i.e. point (2) in Fig.\ref{fig:stress_peak}(a). This brings the system to the onset of melting, see point (2) in Fig.\ref{fig:stress_peak}(b). At this point, the system is not disordered yet and we discuss this structural transition in detail later on. Here, this nearly linear increase in shear stress represents that the system response is nearly elastic at low strains. Further accumulation of the strain melts the crystal and, accordingly, causes a release of stress due to structural relaxation. This brings the system to a strain-independent plateau at higher strains. We can associate this best with the shear-induced disordering transition where the equilibrium melting transition is displaced by the imposed shear rate\cite{ackerson1981shear, holmqvist2005crystallization, butler1995shear}. 

Moreover, including the self-propulsion helps shear by causing a preliminary melting, which appears as a decrease in $\psi_6$ during the relaxation of the system. This fastens the hexagonal-liquid transition and consequently decreases the stress barrier, which is defined by the shear stress peak. Much higher activities ($Pe_a\geq30$) melt the crystal entirely even before imposing the shear rate, thereby annihilating the stress barrier. After that, the colloidal suspension is disordered and its behaviour is liquid-like [in Fig.\ref{fig:stress_peak} snapshot(3)]. In the latter case, the equilibrium melting transition is displaced by self-propulsion and melting is expected for the parameters of the system~\cite{bialke2012crystallization}.

Points (2) in Fig.\ref{fig:stress_peak}(a) and (b) together reveal that the stress peak appears just before the drop of $\psi_6$  below the structural transition, $\psi_6 = 0.45$\cite{bialke2012crystallization}.  
In fact, the hexatic-to-liquid transition is not direct throughout this barrier but rather occurs via a structural rearrangement [Fig.\ref{fig:stress_peak} snapshot(2)]. The zigzag motion of the particles, which emerges by their alternating motion in one layer between filling the grooves of the next layer and hopping up from there\cite{chen1990discontinuous}, temporarily gives a rise in $\psi_4$ [see Fig.\ref{fig:stress_peak} (c)]. We might expect this increase of $\psi_4$ at first sight when we look at the arrangement of particles in the snapshot (2). Additionally, even in the liquid regime below $\psi_6 < 0.45$, where the shear melts the crystal as a whole, some ordered regions show up at times due to this zig-zag motion of the particles in subsequent layers. This is similar to what was observed by Wu et. al.\cite{wu2009melting} in melting passive colloidal suspensions under shear. Although these ordered structures give rise to local increases in the hexagonal order parameter in the passive system below the structural transition line [see Fig.\ref{fig:stress_peak}(b)], the self-propulsion destroys these local ordered structures, resulting in zero $\psi_6$ [see the Supplementary Material, $Movie\_Passive\_Pes20.mp4$].

Furthermore, especially in active cases, we see that the subsequent melting first starts occurring at the least sheared domain of the system and spreads toward the boundaries through the accumulation of interstitial defects as the activity increases. The self-propulsion reveals these interstitial defects at weakly sheared domains and in this way helps the shear to melt the suspension. However, the highly sheared regions of the domain, i.e. regions in proximity to the top and bottom boundaries, are able to temporarily restore their ordered structures and flow as sliding layers for a longer time by resisting the self-propulsion force. Increasing self-propulsion force amplifies the defects to grow and spread toward the boundaries, resulting in the total melting of the system at the end [see the Supplementary Material, $Movie\_Pea20\_ interstitialdefects.mp4$].

\subsection{Motility-induced shear thickening (MIST) and shear thinning}\label{section32}
\noindent We now consider the liquid regime and discuss  the shear stress $ \bigl \langle \sigma_{xy} \bigr \rangle $ with respect to the imposed shear rate $\dot{\gamma}$ [see in Fig.\ref{fig:power_law}(a)], which yields the rheology of the active colloidal suspensions. This time, we average the shear stress data over the time frame corresponding to the strain-independent plateau of the stress. 

The most common rheological model, 
the power law model\cite{saramito2016complex} $\sigma_{xy} = K \dot{\gamma}^n$, is used henceforth to quantify the possible non-Newtonian behaviour of the system. We fit our stress-shear rate data from the simulations with different activities to this model [Fig.\ref{fig:power_law}(b)]. Thereby, we explore the effect of the self-propulsion on the rheological response of the system. 
In this case, we follow up the structural changes by the number of surrounding particles $N_{sp}$, which is the number of particles counted inside the surrounding circle with a radius $r_{surr}/a=3$ for every particle individually [see the top snapshots in Fig.\ref{fig:power_law}]. 

The passive system, initially in the crystalline phase, does not melt at very low shear rates $(Pe_s <2.5)$ and shows solid-like behaviour [Fig.\ref{fig:power_law}(1)] with non-zero yield stress, see in Fig.\ref{fig:power_law}(a). This is in line with what Chen et. al. demonstrated for passive ordered suspensions\cite{chen1990discontinuous}. Here, thermal fluctuations are responsible for the motion of colloids. At this solid regime, we see that turning on the activity diminishes the yield stress. In addition to weak activities, the hexagonal order in the system can be destroyed by more effective shearing, above the so-called critical shear rate\cite{butler1995kinetics, butler1995shear}. On the other hand, activities $Pe_a >30$ eliminate the solid-like behaviour, since the hexatic-to-liquid transition is already reached by activity [see again in Fig.\ref{fig:stress_peak}(b)]. Thereafter, the resulting melt starts behaving as a shear-thinning liquid which is a non-Newtonian behaviour characterized by the slopes lower than 1 in the logarithmic shear stress-shear rate curve, then giving power indexes $n<1$, see point (2) in Fig.\ref{fig:power_law}(b).
Here, the shear force has control over the motion of the particles and yields sliding layers of particles aligned in the direction of shear flow [snapshot 2 in Fig.\ref{fig:power_law}]. Indeed, this layered flow, which is also discussed in Section \ref{section31}, has been previously found as being intrinsic to the shear thinning behaviour of passive Brownian colloidal suspensions\cite{chen1990discontinuous,chen1994rheological, wilemski1991nonequilibrium}. Moreover, the local ordered regions temporarily observed as a result of the zig-zag motion of colloids is known as a shear-induced ordering phenomenon for 2D shear-thinning colloidal suspensions\cite{ackerson1988shear}. Although the shear thinning behaviour is relatively well understood in  Brownian colloidal systems, the effect of self-propulsion has not been addressed. We observe that activities up to $Pe_a \approx 60$  assist only the shear in melting the system, while the resulting melt maintains the same shear-thinning behaviour observed for passive suspensions [see the Supplementary Material, $Movie\_shearthinning.mp4$].

Interestingly, we find that increasing the self-propulsion force further introduces a transition in the rheological response of the suspension. At moderate motilities, $Pe_a \approx 70$, the fluid behaviour is not shear-thinning, but Newtonian, corresponding to the curves (3) with a linear dependence in Fig.\ref{fig:power_law}(a,b). In this regime, self-propulsion degrades the stability of layered flow and the suspension becomes completely disordered, see snapshot (3) in  Fig.\ref{fig:power_law}, [also see the Supplementary Material, $Movie\_Newtonian.mp4$]. Similar to the mechanism discussed in Section \ref{section31}, some of the interstitial defects appeared in shear-thinning liquid start spreading with increasing self-propulsion, being responsible for this transition.

Above $Pe_a \approx 80$, we observe that the self-propulsion starts dominating over shear and reveals structural heterogeneities in the system, leading to transient cluster formation as shown by the snapshot (4) in Fig.\ref{fig:power_law}, [also see the Supplementary Material, $Movie\_shearthickening.mp4$]. In this case, the slope of shear stress- shear rate curves exceeds 1 [see (4) in Fig.\ref{fig:power_law}(b)],
meaning that the suspension reveals a second transition from Newtonian to shear thickening behaviour. Here, we also observe that the colloidal suspension starts showing clusters even before imposing a shear to the system, which is expected for active particles in this parameter regime~\cite{digregorio2018full} [see the Supplementary Material, at the beginning of $Movie\_shearthickening.mp4$ in the absence of shear]. However, we found that these clusters  enable the system to behave as a shear-thickening fluid when the system is forced to flow. Therefore, we refer to this behaviour as {\it motility-induced shear thickening} (MIST) in some analogy to motility-induced phase separation (MIPS). In fact, the cluster formation generating MIST is a precursor of MIPS. Additionally, especially for this behaviour, we observe visually that the number of surrounding particles $N_{sp}$ remarkably differs from region to region in the system [see snapshot (4) in Fig.\ref{fig:power_law}].

The shear thinning-to-thickening transition has been attained for years in passive colloidal suspensions with very high shear rates $(Pe_s>100)$, by underlining similar cluster formations in the shear-thickening fluid due to the hydrodynamic interactions and lubrication forces\cite{bossis1989rheology, hoffman1998explanations, brown2014shear, lorenzo2022brownian}.  In this study, instead, we could reach this transition by keeping the range of shear rate constant at relatively low values, while increasing the activity of colloids, with neglecting the hydrodynamics. In this work, the cluster formation is not triggered by shearing extremely the system, rather is triggered by motility.

\subsection{The effect of particle density}
\begin{figure}[h!]
    \centering
    \includegraphics[width=0.45\textwidth]{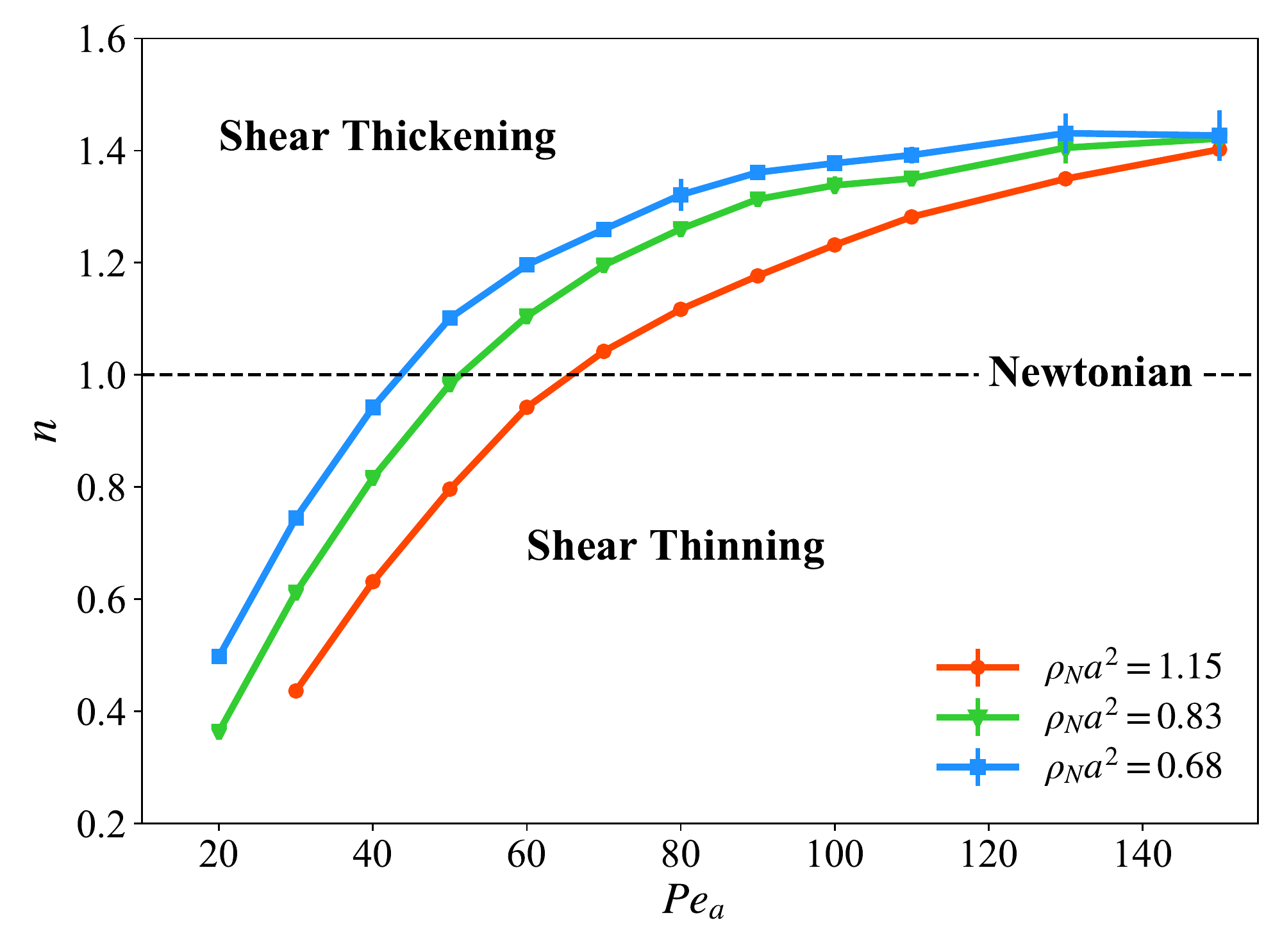}
    \caption{The power index $n$ vs the active Peclet number $Pe_a$ for different particle densities. The horizontal dashed line corresponds to the Newtonian case with a slope of 1 on the log-log scale.}
    \label{fig:n_vs_act}
\end{figure}
Next, we analysed the stress-shear rate curves for the suspensions with two lower particle densities,  semi-dilute suspension $\rho_N a^2=0.83 $, and dilute suspension $\rho_N  a^{2}  =0.68$, respectively, keeping all other parameters the same in the system. Decreasing the number of particles in the computational box gives more space to particles for disordering, compared to the dense suspension system. This, consequently, increases the tendency of the system to be melted. Thus, the solid behaviour is found as limited to very small parameter ranges. Additionally, the dilute suspensions show lower yield stresses compared to the dense case at the same activities. Accordingly, the system starts showing a shear thinning behaviour even with very low self-propulsion forces. At this point, Fig.\ref{fig:n_vs_act} interprets the trend of the power index with increasing self-propulsion for all cases. Here, again, the power indexes are obtained by fitting the shear stress/shear rate data to the power law model for these two other densities, as done for the dense system in Section \ref{section32}. The red line in this figure, which corresponds to the dense suspension case, starts at $Pe_a \approx30$, since the system retains mostly solid-like behaviour below this value, followed by shear-induced disordering with increasing activity.

 Regarding the liquid-like regime, it is evident that the critical self-propulsion for shear thinning to thickening transition shifts to lower activities with decreasing particle density. In other words, dilute suspensions can thickens easier than dense ones. The underlying reason for this effect can be intuitively related to clustering. Keeping the particles very close to each other, as in a dense system, obviously pronounces their collisions. Therefore, possible cluster formation becomes difficult in the system. Although some clusters are formed with these activities, they are not stable enough, such that they can be damaged by any strike from other particles around them. However, increasing the self-propulsion further promotes clustering and finally brings the system into a thickening regime. As we discussed previously, activity triggers the colloidal suspension to behave as a shear-thickening fluid in this study, not strong shearing.  
 \begin{figure}[h!]
    \centering
    \includegraphics[width=0.45\textwidth]{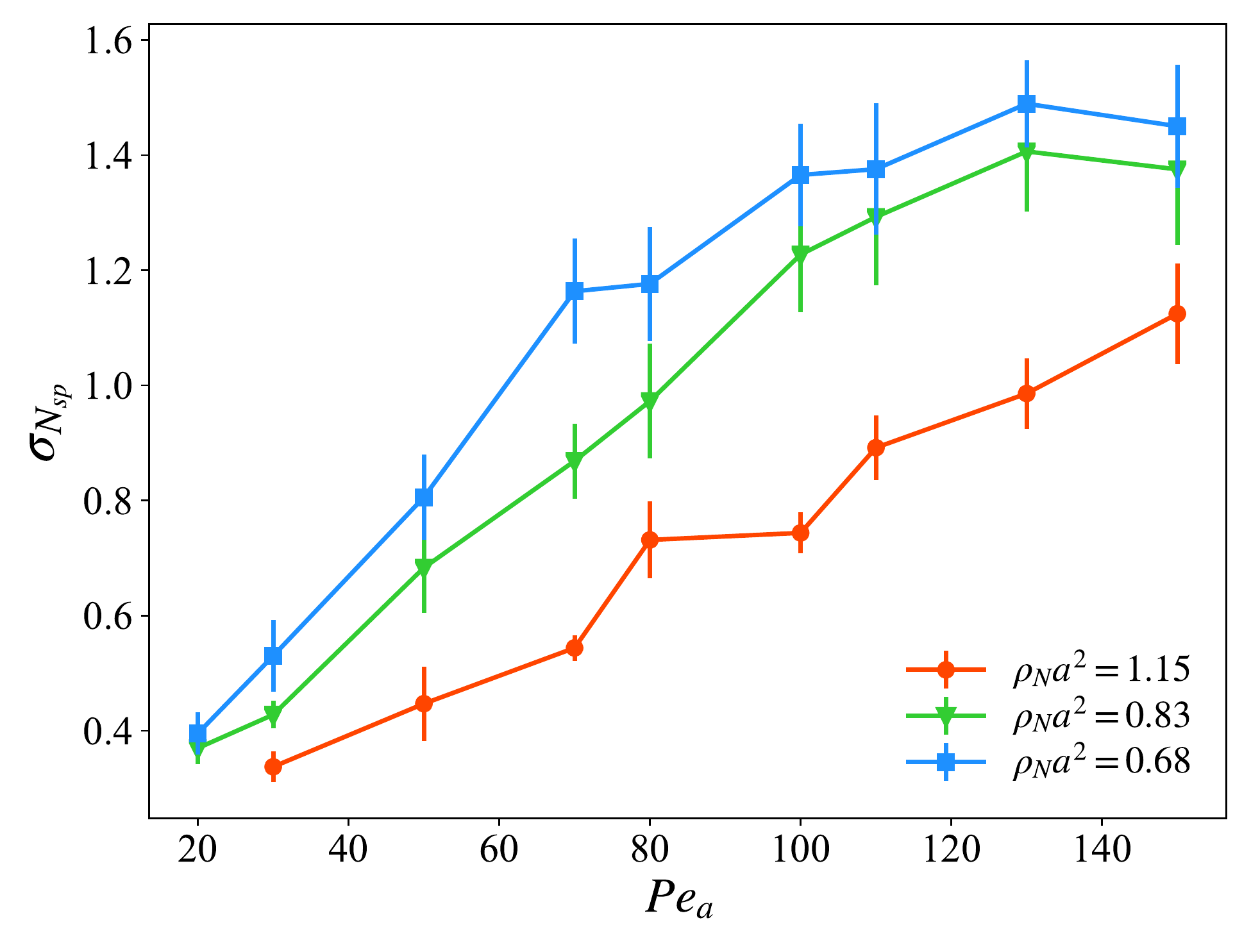}
    \caption{Standard deviations in the number of surrounding particles, $\sigma_{N_{sp}}$ with increasing self-propulsion $Pe_a$ for different particle densities, $\rho_N a^2=1.15$ (circle), $\rho_N a^2=0.83$ (triangle) and $\rho_N a^2=0.68$ (square) at $Pe_s=20$.}
    \label{fig:Nsp}
\end{figure}
 On the other hand, active colloids in a dilute environment can constitute more stable clusters even with low self-propulsion due to fewer repulsive collisions coming from particles around the clusters, accordingly undergoing this thinning-to-thickening transition earlier [see Fig.\ref{fig:n_vs_act}].

This scenario is also confirmed by Fig.\ref{fig:Nsp}, where we report the standard deviations $\sigma_{N_{sp}}= \sqrt{\langle N_{sp}^2 \rangle - \langle N_{sp}\rangle ^2 }$ of the number of surrounding particles from the mean $\langle N_{sp} \rangle$  for the three different densities.

Typically there are more fluctuations in the surrounding particle density for dilute systems, also for weak self-propulsion forces. However, here the activity enhances dramatically (i) the local density inhomogeneities, as revealed by the strong increase of $\sigma_{N_{sp}}$ with $Pe_a$, and (ii) the gap in $\sigma_{N_{sp}}$ between dilute and dense systems. The motility-induced local density inhomogeneities correspond to pre-clustering which provides evidence that this is the underlying reason for the shear-thickening behaviour consistent with snapshot (4) in Fig.\ref{fig:power_law}. Since this pre-clustering is more pronounced for dilute systems, the MIST transition occurs earlier in these systems confirming our previous intuitive explanation.

\section{Conclusions}
In this work, we have explored the effect of self-propulsion on the rheological response of the dense colloidal suspensions under steady shear by using Brownian dynamics simulations. 
First, the solid-to-liquid transition of the suspension was characterized, showing that activity helps to melt the systems, as shown before in literature\cite{bialke2012crystallization,li2015effects}.
When melting, the self-propulsion of the colloids reduces the stress barrier that the system has to overcome in order to transition from the solid to the liquid state.
Self-propulsion is not only responsible for assisting the shear in melting the colloidal suspension but also introduces a transition in the rheology of the melted suspension. Depending on the activity of colloids in the suspension, different fluid behaviours are found: shear thinning, Newtonian and shear-thickening. The underlying reason behind this response is a well-known dynamical mechanism of active colloidal particles: activity-induced clustering. When the self-propulsion is sufficiently strong to prevail against the shear, or even before shearing, a cluster formation is  observed. The existence of these clusters creates additional resistance to flow when the suspension is exposed to shear, causing it to behave as a shear-thickening fluid. We referred to this as motility-induced shear-thickening in this article. 

The shear-thinning to thickening transition reported here is different from the transition observed in passive colloidal systems. 
Passive colloids tend to stick together by hydrodynamic and lubrication forces at high shear rates\cite{wagner2009shear, brown2014shear}, while active colloids cluster due to their motility (akin to motility-induced phase separation). 
The simulations presented here were performed in the absence of hydrodynamic interactions. Including hydrodynamics and lubrication forces would be an interesting future avenue since hydrodynamics suppress the tendency for active particles to form clusters~\cite{navarro2015clustering,theers2018clustering,matas2014hydrodynamic}, while on the other hand, hydrodynamics lead to clusters of sheared passive colloids. 
Further, it would be interesting to test the response in a three-dimensional set-up or to include inertia to the active particles motion~\cite{wagner2019response, lowen2020inertial}. 
Experimentally, the shear thinning to thickening transition could be tested using active colloids in a "washing-machine" set up~\cite{williams2022rheology}. 
Finally, it would be interesting to extend our simulation to particles of more complex shape than spheres such as active polymers or active filaments~\cite{kaiser2015does,winkler2020physics},
where new rheological behaviour due to active entanglements can be expected in dense solutions.

\section*{Conflicts of interest}
There are no conflicts to declare.\

\section*{Acknowledgements}
We thank J\"urgen Horbach for helpful discussions. The work of A.G.B. was supported within the EU MSCA-ITN ActiveMatter (Proposal No. 812780).

\section*{Supplementary Material}

$Movie\_ Passive\_ Pes20.mp4 $:
The movie of the simulation with passive colloids at $Pe_s=20$. The snapshots in Fig. 3 of the main paper are taken from this movie. The system starts from the hexagonal crystal phase and the subsequent shearing promotes the sliding layers throughout the melting. The local ordered regions emergent due to the zig-zag motion of particles in these sliding layers are distinguishable at times. 
\\
$Movie\_Pea20\_ interstitialdefects.mp4 $: The movie for the simulation with active colloids ($Pe_a=20$) under shearing ($Pe_s=20)$, shows that the self-propulsion assists the shear in disordering the system by promoting the interstitial defects. These defects start appearing at the least sheared region of the system, and gradually spread toward the system boundaries with increasing self-propulsion. However, we show only their emergencies here for the suspension system at a constant self-propulsion.  
\\
$Movie\_shearthinning.mp4 $: The movie corresponds to the shear-thinning regime, i.e. $Pe_a=30$. The colloidal suspension here is melted entirely by self-propulsion. After imposing the shear, the particles start to reorganise and cause a layered flow, which is a peculiar behaviuor of shear-thinning fluids. Some locally ordered domains appear at times due to the zig-zag motion of particles under shear flow. 
\\
$Movie\_Newtonain.mp4 $: The movie corresponds to the Newtonian regime, i.e. $Pe_a=70$. Colloids are homogeneously disordered by self-propulsion. 
\\
$Movie\_shearthickening.mp4 $: The movie corresponds to the shear-thickening regime, i.e. $Pe_a=150$, the so-called the motility-induced shear thickening (MIST). Self-propulsion induces structural heterogeneity and clustering in the system.
\bibliography{references}

\end{document}